\documentclass[english,twocolumn,superscriptaddress,
  longbibliography]{revtex4-1}

\usepackage{xcolor}
\usepackage{graphicx}
\usepackage{amsmath}
\usepackage{amsfonts}
\usepackage{color}
\usepackage{babel}
\usepackage{url}
\usepackage[breaklinks]{hyperref}
\usepackage{tikz}
\usepackage{CJKutf8}
\usepackage[title]{appendix}

\definecolor{lime}{HTML}{A6CE39}
\DeclareRobustCommand{\orcidicon}{
	\begin{tikzpicture}
	\draw[lime, fill=lime] (0,0) 
	circle [radius=0.16] 
	node[white] {{\fontfamily{qag}\selectfont \tiny ID}};
	\draw[white, fill=white] (-0.0625,0.095) 
	circle [radius=0.007];
	\end{tikzpicture}
	\hspace{-2mm}
}

\foreach \x in {A, ..., Z}{\expandafter\xdef\csname
  orcid\x\endcsname{\noexpand\href{https://orcid.org/\csname
      orcidauthor\x\endcsname} {\noexpand\orcidicon}} }

\begin{document}

\title{Approximate Entropy Analysis for Nonlinear Beam Dynamics}

\author{Yongjun Li\orcidA{}}\thanks{email: yli@bnl.gov}
\affiliation{Brookhaven National Laboratory, Upton 11973, New York, USA}
 
\begin{abstract}
  In this paper, we apply approximate entropy (ApEn) analysis to the
  nonlinear beam dynamics in circular accelerators. Due to the presence of
  strong nonlinear magnets, chaos of beam motion gradually increases with
  amplitude. Such chaos can be quantitatively characterized with ApEn of
  beam turn-by-turn readings. Then ApEn, as a chaos indicator, can be used
  for nonlinear lattice optimization and analysis.
\end{abstract}
 
\maketitle

\section{\label{sect:intro}introduction}

  For circular particle accelerators, the nonlinearities of beam dynamics
  confine long-term motions to be stable only within a limited region in
  6-dimensional phase space, namely, dynamic aperture
  (DA)~\cite{chao2023hb}. Even within DA, particle motions could still be
  chaotic. It is commonly believed that, for a given magnetic lattice,
  through suppressing chaos, one can enlarge its DA and local momentum
  acceptance (LMA), and also enhance its robustness to errors. Therefore,
  various chaos indicators have been adopted to characterize the
  nonlinearities of beam motions~\cite{Bazzani:2023hbb}, such as the
  Lyapunov exponent (LE)~\cite{wolf1985determining, schmidt1991comparison,
    Habib:1995}, frequency map analysis
  (FMA)~\cite{laskar1999introduction}, forward-reversal integration
  (FRI)~\cite{panichi2017reversibility, li2021fast, borland2000elegant},
  data-driven chaos indicator~\cite{li2022data}, fluctuation of
  approximate invariant~\cite{li2021design}, etc. In this paper we apply
  approximate entropy (ApEn) to analyze the nonlinear beam dynamics.

  The concept of entropy has its origins in classical physics under
  the second law of thermodynamics. In the context of nonlinear
  dynamics, information entropy is central in quantifying the degree of
  uncertainty or information gain, and is therefore widely used to
  explain complex nonlinear behavior in real-world systems. Among many
  entropy analyses, ApEn conception was initially developed by
  Pincus~\cite{pincus1991approximate} to analyze medical data, such as
  heart rate, and later spread its applications in many other fields,
  such as finance~\cite{pincus2004irregularity}, and nonlinear
  dynamics~\cite{volos2020nonlinear}, etc. It is a technique used to
  quantify the amount of regularity and the unpredictability of
  fluctuations particular for short and noise time-series
  data.
  
  Due to the presence of strong nonlinear magnetic fields in circular
  accelerators, the chaos of beam motions gradually increase with
  amplitude. Such chaos is often visualized with the Poincar\'{e} map,
  i.e., the intersection of a periodic orbit projected in a certain
  lower-dimensional subspace, usually a 2-dimensional coordinate-momentum
  phase space. Experimentally it can also be observed from turn-by-turn
  (TBT) data after beam is excited. ApEn can quantitatively characterize
  the chaos of circulating beam TBT readings. Based on that, we can
  determine if magnetic lattices are well configured or not. Thus, ApEn,
  as a chaos indicator, can be used for nonlinear lattice optimization at
  design stage, or online beam-based optimization as
  well~\cite{huang2018robust} if BPMs have TBT resolution.

  The remainder of this paper is outlined as follows:
  Sect.~\ref{sect:apen} reviews the definition of ApEn and explains its
  principle briefly. Then ApEn analysis is applied to a H\'enon map as a
  proof-of-principle in Sect.~\ref{sect:henon}. In
  Sect.~\ref{sect:lattice}, ApEns observed in the transverse $x$--$y$
  plane are used as minimization objectives to optimize the National
  Synchrotron Light Source II (NSLS-II) nonlinear lattice. In
  Sect.~\ref{sect:info}, we implement a detailed ApEn analysis for an
  elite candidates selected from the previous optimization, and interpret
  physics information being conveyed by such analysis. A summary is given
  in Sect.~\ref{sect:summary}.

\section{\label{sect:apen}Approximate entropy}

  According to Pincus~\cite{pincus1991approximate}, ApEn is defined
  as: Two input parameters, $m$ and $r$, must be fixed to compute
  ApEn, $m$ is the ``length" of compared runs, and r is effectively a
  filter. Given N data point $u = {u(1), u(2), \dots, u(N)}$ form
  vector sequences $x(1)$ through $x(N-m+1)$, with $m \le N$ define by
  $x(i) = {u(i), u(i+1),\dots, u(i+m-1)}$. These vectors represent $m$
  consecutive $u$ values, starting with the $i^{th}$ point.  Define
  the distance $d[x(i),x(j)]$ between $x(i)$ and $x(j)$ as the maximum
  difference in their respective scalar components, $d[x(i),x(j)]=
  \max_{k=1,2,\dots,m}(|u(i+k-1)-u(j+k-1)|)$. Then we calculate the
  value $C^m_i(r) = (\text{number of}\,j \le N-m+1\, \text{such
    that}\,d[x(i),x(j)] \le r)/(N-m+1)$. The numerator of $C^m_i$
  counts, within the tolerance $r$, the number of blocks of
  consecutive values of length $m$ which are similar to the pattern of
  window length $m$. Then the $C^m_i$'s measure the regularity, or
  frequency, of patterns similar to this window. Define
  \begin{equation}
    \varphi^m(r) = \frac{1}{N-m+1} \sum_{i=1}^{N-m+1}\log C^m_i(r),
  \end{equation}
  where $\log$ is the natural logarithm. The ApEn is
  defined as
  \begin{equation}
    \text{ApEn}(m,r,N)(u) =
    \lim_{N\rightarrow\infty}\varphi^m(r) - \varphi^{m+1}(r),
  \end{equation}
  which is a measure of system complexity. Fundamentally, ApEn
  measures the conditional probability that nearby pattern runs remain
  close in the next incremental comparison. A positive ApEn usually
  indicates a chaos~\cite{eckmann1985ergodic} . The value of $N$, the
  number of input data points for ApEn computations, is typically
  between 75 and 5,000. Based on the calculations that included both
  theoretical analysis and clinical applications, Pincus concluded
  that, for $m=1,2$, values of filter $r$ between 0.1 to 0.25 standard
  deviation of the $u(i)$ data can produce good statistical
  validity~\cite{pincus1995approximate}. We found that such settings
  are also applicable to nonlinear beam dynamics for circular
  accelerators.
  
  ApEn can be computed directly based on the above definition. It requires
  execution time analogous to the square of the size of the input
  signal. Fast algorithms, such as~\cite{manis2008fast}, were proposed to
  speed up its computation. In the meantime, well-developed and documented
  computation packages, such as EntropuHub~\cite{flood2021entropyhub}
  (used in this paper), are also available.

\section{\label{sect:henon}Application to H\'{e}non map}

\begin{figure}[htpb]
     \includegraphics[width=0.95\columnwidth]{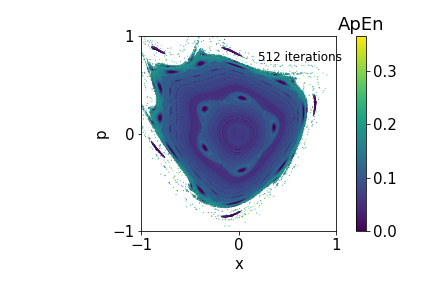}
     \includegraphics[width=0.95\columnwidth]{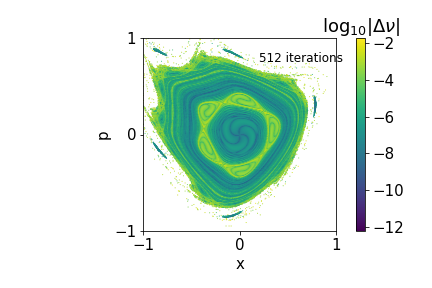}
     \includegraphics[width=0.95\columnwidth]{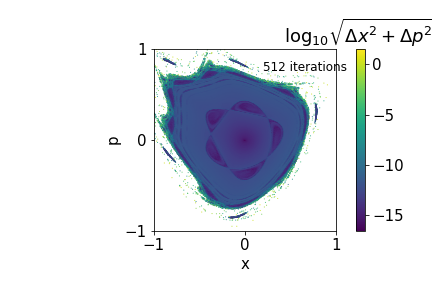}
     \caption{\label{fig:henon}(Colored) Comparison of the Apen (top), FMA
       (middle) and FRI (bottom) analyses for a H\'enon map with 512
       iterative runs. The color-maps represent the ApEn, tune diffusion
       and distance between forward and reversal trajectories respectively
       at the locations of their initial conditions. The blank area
       represents unbounded trajectories.}
  \end{figure}

  In this section, ApEn analysis is applied to a 1-dimensional H\'enon's
  quadratic area-preserving map~\cite{henon1969numerical},
  \begin{equation}
    \left(\begin{array}{c}
      x\\
      p
    \end{array}\right)_{n}=\left(\begin{array}{cc}
      \cos\mu & \sin\mu\\
      -\sin\mu & \cos\mu
    \end{array}\right)\left(\begin{array}{c}
      x\\
      p-\lambda x^{2}
    \end{array}\right)_{n-1}.
  \end{equation}
  From the view of beam dynamics, this discrete map represents a thin-lens
  sextupole kick followed by a linear phase space rotation at a phase
  advance $\mu=2\pi\nu$. Here, sextupole strength $\lambda=1$, and the
  linear tune is chosen to be $\nu=0.205$ to observe the $5^{th}$-order
  resonance at certain amplitudes. The map is iterated for 512 runs, then
  the ApEns of $x$-coordinates are computed for each initial condition in
  the phase space $x$-$p$ as illustrated in Fig.~\ref{fig:henon}. Its ApEn
  (chaos) increase gradually with the initial condition's amplitude
  $\sqrt{x^2+p^2}$, except while crossing stable resonances. The contours
  of FMA and FRI analyses are also computed for the purpose of comparison.

\begin{figure}[htpb]
     \includegraphics[width=0.9\columnwidth]{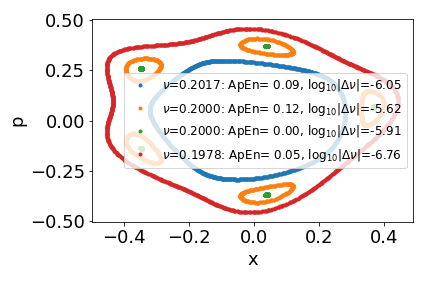}
     \caption{\label{fig:ApEnvsFMA}(Colored) Trajectories of the
       H\'{e}non map in the $x$--$p$ phase space while crossing the
       $5^{th}$ order resonance stably. Once tune is sufficiently
       close to 0.2, and trajectories are trapped by elliptic fixed
       points, motions in the time domain become highly regular with a
       low periodicity 5, then near-zero ApEns are observed.}
  \end{figure}

  Note that, while crossing stable resonances, the behavior of FMA differs
  from ApEn and FRI analyses. Fig.~\ref{fig:ApEnvsFMA} shows four
  trajectories in the Poinc\'{e}re section $x$--$p$, while a $5^{th}$
  order resonance is being crossed driven by the amplitude dependent
  detuning. When the tune is sufficiently close to 0.2, and some
  trajectories are trapped by elliptic fixed points, five isolated islands
  are formed and then gradually merge to five fixed points. Around those
  stable fixed points, the fundamental tune diffusions are relatively
  large (at the order of $10^{-5\sim6}$). However, the dimension of
  islands is relatively small, then time-series data composed of $x$ (or
  $p$) is highly regular with a low periodicity, which yields near zero
  ApEn. Similar behaviors are observed in the difference of FRIs as
  well~\cite{li2021fast}. It could be explained as, for an on-resonance
  but stable time-series data, its regularity is higher in the time domain
  than in the frequency domain. Another observation from
  Fig.~\ref{fig:henon} is that, the FMA and ApEn could provide more rich
  information (having more fine structures in their chaos maps) than the
  FRI analysis. It means they could be more sensitive to the variations of
  chaos.

\section{\label{sect:lattice}Application to nonlinear lattice optimization}

  In this section, we use ApEn as chaos indicator to optimize the
  nonlinear lattice for the NSLS-II storage ring. In a linearly stable
  lattice, the motion of particle, seen by a BPM at a certain location, is
  a periodical time-series oscillating with a fixed frequency, known as
  the linear tune. Nonlinear magnets, such as sextupoles for chromaticity
  correction, can perturb regular motions. Thus, signals seen by the BPM
  now have fluctuations on top of regular motions. The ApEn of TBT BPM
  readings reflects the likelihood that similar patterns of the TBT
  readings will not be followed by additional similar readings. Given a
  nonlinear lattice, if the ApEns of TBT readings are low, the beam motion
  is less chaotic, and vice versa. By minimizing the ApEn of different
  trajectories through tuning nonlinear knobs, the lattice could be
  optimized. Usually, ApEn analysis can let chaos be visible from a short
  time-series with only several tens of data. It means that, only
  short-term TBT data is needed to drive optimizer. At the early stage of
  lattice design, such low computation cost can efficiently narrow down
  the search range by ruling out bad candidates.

\begin{figure}[htpb]
    \centering \includegraphics[width=\columnwidth]{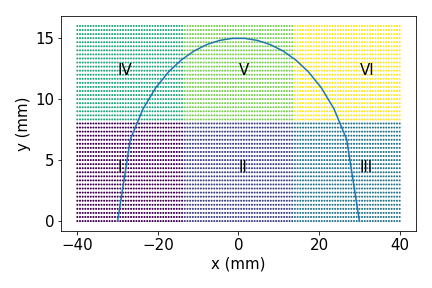}
    \caption{\label{fig:zone} (Colored) Dividing the region of interest
      (RoI) into $2\times3=6$ zones in the $x$--$y$ plane at the
      observation. In each zone, multiple initial conditions (represented
      with same-colored dots) are uniformly populated. The optimization
      objectives are the averaged ApEns of all initial conditions within
      each individual zone. The solid line is the desired DA profile for
      the NSLS-II storage ring.}
  \end{figure}

  Next, we explain the detailed optimization implementations using the
  NSLS-II storage ring as an example. From a certain longitudinal
  observation location (the injection point in this example), multiple
  initial conditions are uniformly populated within a Region of Interest
  (RoI) with a transverse dimension $x\in[-40,40]\, y\in[0,15]\,mm$
  (Fig.~\ref{fig:zone}). The RoI should be chosen slightly larger than the
  desired dynamic aperture. Particle trajectories are simulated with a
  symplectic integrator~\cite{Yoshida:1990} implemented in the code
  \textsc{elegant}~\cite{borland2000elegant}. Each trajectory is simulated
  for $N=256$ turns if survived, and TBT data at the observation point is
  recorded, then used to compute its ApEn.

  The goal of optimization is minimizing the ApEns for all initial
  conditions within the RoI. It is neither practical nor necessary to
  minimize every initial condition simultaneously, therefore, the RoI is
  divided into several zones as shown in Fig.~\ref{fig:zone}. For each
  zone, two objectives are the horizontal and vertical ApEns averaged over
  all initial conditions if survived after $N=256$ turns.  By tuning the
  sextupole knobs, we attempt to minimize each zone's ApEns simultaneously
  to suppress the overall chaos inside the whole RoI. The reason of
  choosing $N=256$ turns is that such data-sets are already long enough to
  measure the chaos with ApEn.  In the meantime, the computation cost
  needed for particle tracking simulation is relatively low. Usually, most
  light source rings are composed of multiple identical cells, therefore,
  we can even use one cell to optimize the DA for the ideal error-free
  lattice to narrow down the search range quickly. Then in the next stage,
  a full ring lattice including various imperfections is used to search
  robust solutions within a narrow range.
  
  ApEn need be computed in the horizontal and vertical planes
  respectively, and they are usually at different scales and need to be
  minimized separately. Then the number of optimization objectives is two
  times the number of zones. The tuning knobs in this example are the
  normalized gradients $K_2$ of six harmonic sextupole families. The range
  of $K_2$ is within $[-40,40]\,(m^{-3})$ limited by their power supply
  capacities and magnetic saturation. These harmonic sextupoles doesn't
  contribute to the linear chromaticity, but can compensate the geometric
  and chromatic optics abbreviations generated by chromaticity correction
  sextupoles.

\begin{figure}[htpb]
     \includegraphics[width=0.9\columnwidth]{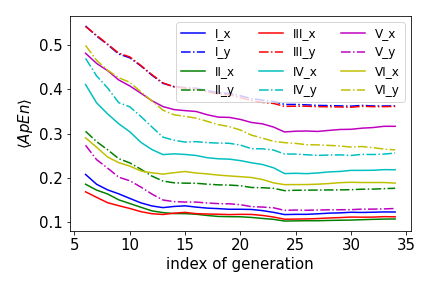}
     \caption{\label{fig:moga}(Colored) Convergence of averaged ApEns in
       the genetic algorithm optimization.  Solid lines stand for the
       horizontal plane, and dashed-dotted ones for the vertical
       plane. Lines with a same color are from the same zone. }
  \end{figure}

\begin{figure}[htpb]
     \includegraphics[width=0.8\columnwidth]{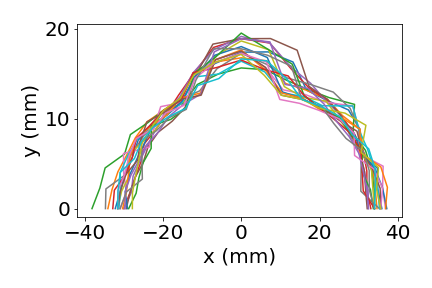}
     \includegraphics[width=0.8\columnwidth]{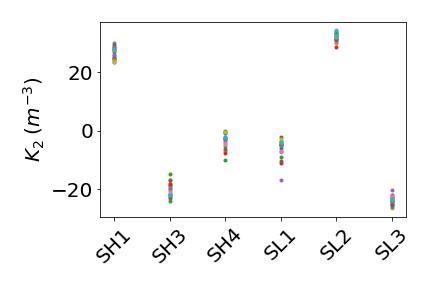}
     \caption{\label{fig:top3da}(Colored) Top: on-momentum DA profiles for
       20 elites selected from the last generation of the genetic
       optimization. Bottom: distribution of six sextupole knob
       configurations of these elites. They converge to some narrow ranges
       as well.}
  \end{figure}

\begin{figure}[htpb]
     \includegraphics[width=0.9\columnwidth]{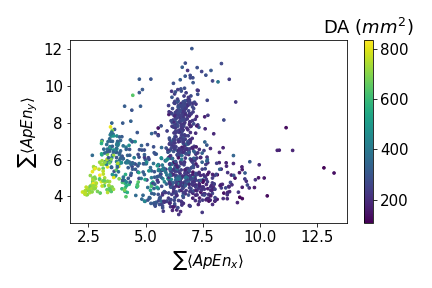}
     \caption{\label{fig:correlation}(Colored) Correlation between ApEn
       and DA. The horizontal and vertical axes are the sums of six zone's
       averaged ApEns. Each dot represents one candidates in the $30^{th}$
       generation of the genetic optimization, colored with its area of
       on-momentum DA. The correlation confirms that having small ApEns in
       both the horizontal and vertical planes is essential to enlarge the
       NSLS-II ring DA.}
  \end{figure}

  This multi-objective optimization is solved with the widely used genetic
  algorithm~\cite{deb2002fast, yang2011moga,li2018ml}. A small population
  with 1,000 candidates evolves more than 30 generations, a good
  convergence of the average ApEn has been reached
  (Fig.~\ref{fig:moga}). The DAs of all candidates in the $30^{th}$
  generation are calculated for picking up some elite candidates among
  them. The on-momentum DA profiles of top 20 elites are illustrated in
  the top subplot of Fig.~\ref{fig:top3da}. It is interesting to observe
  that the distributions of their six knobs (sextupole gradients $K_2$)
  also converge to some small ranges as shown in the bottom subplot. The
  correlation between the average ApEns and the area of DAs is illustrated
  in Fig.~\ref{fig:correlation}, which confirms that suppressing the ApEns
  in both the horizontal and vertical planes is essential in enlarging DA.

\begin{figure}[htpb]
     \includegraphics[width=0.9\columnwidth]{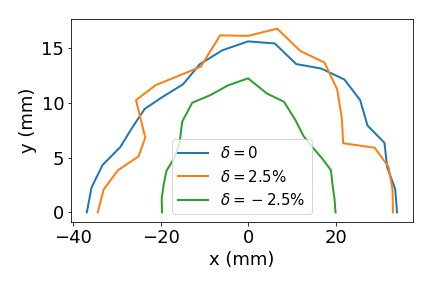}
     \caption{\label{fig:bestda}(Colored) On- and two off-momentum DAs for
       a selected elite candidate.}
  \end{figure}

  In designing a nonlinear lattice, the DA and local momentum acceptance
  (LMA) must be considered simultaneously to satisfy the requirements on
  the injection efficiency and beam lifetime~\cite{Borland:2015}. A same
  strategy as ref.~\cite{li2021fast} is used to include some off-momentum
  DAs as optimization objectives. In this example, on $\delta=\pm2.5\%$
  off-momentum planes, ApEns are added as the objectives, which are
  evaluated in the same way as Fig.~\ref{fig:zone}. Considering
  off-momentum DAs could be smaller than the on-momentum one, slightly
  tight ROIs can be used. The on- and two off-momentum ($\delta=\pm2.5\%$)
  DAs for a selected elite candidate are illustrated in
  Fig.~\ref{fig:bestda}, which should be sufficient to achieve
  high-efficient injections and 3 hours Touschek beam
  lifetime~\cite{bernardini1963lifetime} by comparing with our current
  operation lattice.

\section{\label{sect:info} Information learned from ApEn}

\begin{figure}[htpb]
     \includegraphics[width=0.98\columnwidth]{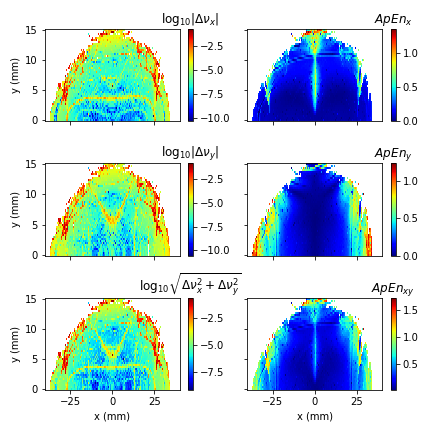}
     \caption{\label{fig:fma_apen_xy}(Colored) FMA and ApEn analyses for a
       selected elite candidate. Left column: Tune diffusion maps of FMA
       observed in the horizontal, vertical and both planes; Right column:
       ApEn maps observed in the horizontal, vertical and both planes.}
  \end{figure}

  Like FMA or FRI etc., ApEn can also provide detailed chaos information
  for a given nonlinear lattice, such as the strength and location of
  resonances, and robustness to errors. Below we use one of elite
  candidates (the same one as shown in Fig.~\ref{fig:bestda}) as an
  example to implement a detailed ApEn analysis. To achieve a more
  accurate result, for each initial condition, a $N=1024$ long TBT data is
  obtained with the code \textsc{elegant}. In the meantime, high density
  initial conditions are populated to produce a high-resolution DA profile
  to identify resonance lines. As already observed in Fig.~\ref{fig:moga},
  the horizontal and vertical ApEns are at different scales. Therefore,
  three ApEn profiles are provided in Fig.~\ref{fig:fma_apen_xy}: two
  separated ApEn maps observed in either the horizontal or vertical plane
  solely, and one weighted map obtained by adding them after normalizing
  with their maxima,
  \begin{equation}
    \text{ApEn}_{x,y} = \frac{\text{ApEn}_x}{
      \max{\text{ApEn}_x}}+\frac{\text{ApEn}_y}{\max{\text{ApEn}_y}},
  \end{equation}
  where $\max{\text{ApEn}}_{x,y}$ are the maxima in each planes
  respectively.

\begin{figure}[htpb]
     \includegraphics[width=0.9\columnwidth]{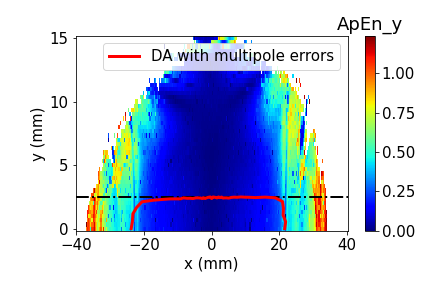}
     \caption{\label{fig:daerror}(Colored) Reduction of DA when a vertical
       physical aperture (dash-dot line) and multipole errors are
       imposed. The horizontal aperture is lost in the area where the
       vertical ApEn is high. The survived DA (red solid line) confirms
       that having low ApEn is essential to obtain robust solutions.}
  \end{figure}

  In the horizontal $ApEn_x$ map, besides visible resonances around
  $x=\pm20\,mm$, strong chaos also appears in the vicinity of y-axis
  ($x=0$), particularly when $y\ge10\,mm$. While in the vertical plane,
  strong chaos shows up when $x\ge|20|\,mm$. Based on this observation, we
  can conclude that, a strong nonlinear cross-talking must exist between
  two transverse planes. Such cross-talking drives the vertical motion to
  be chaotic at large horizontal amplitudes and vice versa. In the
  vicinity of y-axis, small amplitude horizontal TBT data is polluted by
  the coupling from the vertical plane, thus its signal-noise-ratio is
  low, which only cause some visual chaos there. The real problem is that,
  such cross-talking would cause horizontal DA to reduce significantly in
  the presence of vertical physical apertures and errors. We use the
  $ApEn_y$ (see Fig.~\ref{fig:fma_apen_xy}'s middle-right subplot) to
  further illustrate this consequence. When particle's horizontal
  amplitude excesses $20\,mm$, although the horizontal motions remain
  regular (shown with ``cold'' colors in the top-right subplot), their
  vertical motions become chaotic dramatically and should have large
  fluctuations while observing their TBT data. Particles can still survive
  in an open space. However, once small vertical physical apertures and
  errors are in place, they can be scraped. Such cross-talking has a
  practical effect for every light source rings, in which in-vacuum
  undulators (IVU) usually have a few millimeters vertical apertures, and
  cause a significant reduction of DA in the horizontal plane. Note that
  most of light source rings need a sufficient horizontal DA for
  injection. As shown in Fig.~\ref{fig:daerror}, after a $\pm2.5\,mm$
  vertical aperture is imposed and the existing NSLS-II magnet systematic
  multipole errors are included, the horizontal DA immediately reduces to
  the area with low $ApEn_y$. If only the area of DA in free space is used
  as the objective of optimization, some candidates might be estimated
  optimistically. Previously, such difficulty has been overcome by
  including physical apertures and errors into the tracking simulation and
  optimization~\cite{Borland:2015}, but it greatly slows down the speed of
  optimization. One of the benefits of using ApEn as the chaos indicator
  is that, even with error-free magnet models, the robustness of lattice
  to errors and physical apertures could be visible and under
  consideration to a certain extent. Therefore, ApEn could be particularly
  useful in the early stage of lattice optimization to narrow down the
  search range for robust solutions, even with lack of information on
  magnet errors.

  The standard FMA analyses are also illustrated in the left column of
  Fig.~\ref{fig:fma_apen_xy} for comparison. Three tune diffusion maps (in
  the horizontal, vertical and both planes respectively) gradually
  increase with their initial amplitude, but there is no obvious jump
  between the regular and chaotic regime as shown in the subplot of
  $ApEn_y$ profile. However, FMA does provide a much clear view on
  resonances.
      
\section{\label{sect:summary}Summary}
  
  We introduce the ApEn concept to analyze the chaos of nonlinear beam
  dynamics. Using the TBT data observed by BPMs, nonlinear lattice
  configurations can be optimized through suppressing their ApEns. Two
  advantages of using ApEn are its low computational demand and robustness
  to noise~\cite{pincus1991approximate}. ApEn can be designed to work for
  small data samples. It might be particularly useful in online beam-based
  optimization for electron storage rings.  Because, valid TBT data can
  last only a few hundreds, or even a few tens turns due to strong
  radiation damping and beam decoherence. another advantage is, when data
  is noisy, the ApEn measure can be compared to the noise level in the
  data to determine what quality of true information may be present in the
  data. These advantages might make ApEn be a suitable objective for not
  only lattice design but also online optimization.
  
  In the meantime, some limitations also exist in the ApEn
  analysis~\cite{richman2000physiological}. The algorithm counts each
  sequence as matching itself to avoid the occurrence of $\log(0)$ in the
  calculations. This step might introduce bias in ApEn, which causes ApEn
  to have two poor properties in practice: ApEn is heavily dependent on
  the record length and is uniformly lower than expected for short
  records. The ApEn analysis sometimes lacks relative consistency. That
  is, if ApEn of one data set is lower than that of another, it should,
  but does not, remain lower for all conditions tested. Particularly, in
  the vicinity of low order resonances, although TBT readings are regular
  due to a low periodicity, such motions are vulnerable to
  errors. Therefore, it might be better to use ApEn along with other chaos
  maps, such as FMA and FRI, to identify the chaos from different aspects
  in the nonlinear beam dynamics. All these chaos maps share same
  times-series (TBT data) and can be implemented in parallel.

  Beside ApEn, many other entropy algorithms, such as Kolmogorov-Sinai
  entropy~\cite{shiryayev1993new,sinai2009kolmogorov}, sample
  entropy~\cite{richman2000physiological}, Fuzzy
  entropy~\cite{de1993definition}, etc., are also available for analyzing
  nonlinear systems. Each algorithm has its pros and cons in quantifying
  chaos. Some further exploration on applying entropy analysis to beam
  dynamics might be interesting and fruitful.
  
\begin{acknowledgments}
  This research used resources of the National Synchrotron Light Source
  II, a U.S. Department of Energy (DOE) Office of Science User Facility
  operated for the DOE Office of Science by Brookhaven National Laboratory
  under Contract No. DE-SC0012704.
\end{acknowledgments}

\bibliography{apen.bib}

\end{document}